\documentclass[aps,prl,twocolumn,floatfix,citeautoscript,hyperref=pdftex]{revtex4-1}
\usepackage{graphicx}
\usepackage{dcolumn}
\usepackage{bm}
\usepackage{color}
\usepackage{amsmath}
\usepackage{amssymb}
\newcommand {\beq} {\begin{equation}}
\newcommand {\eeq} {\end{equation}}
\newcommand {\bqa} {\begin{eqnarray}}
\newcommand {\eqa} {\end{eqnarray}}

\usepackage[colorlinks=true]{hyperref}

\begin{document}

\title{Zero-field finite-momentum and field-induced superconductivity in altermagnets}

\author{Debmalya Chakraborty}
\affiliation{Department of Physics and Astronomy, Uppsala University, Box 516, S-751 20 Uppsala, Sweden}

\author{Annica M. Black-Schaffer}
\affiliation{Department of Physics and Astronomy, Uppsala University, Box 516, S-751 20 Uppsala, Sweden}

\begin{abstract}

We explore the possibilities for spin-singlet superconductivity in newly discovered altermagnets. Investigating $d$-wave altermagnets, we show that finite-momentum superconductivity can easily emerge in altermagnets even though they have no net magnetization, when the superconducting order parameter also has $d$-wave symmetry with nodes coinciding with the altermagnet nodes. Additionally, we find a rich phase diagram when both altermagnetism and an external magnetic field are considered, including superconductivity appearing at high magnetic fields from a parent zero-field normal state.
\end{abstract}

\maketitle

Recently discovered altermagnetism \cite{Smejkal20,Mazin21,Smejkal22a,Feng22,Hayami19,Hayami20,Gonzalez23,Bai23} has opened up a new field of research in condensed matter physics \cite{Smejkal22} by introducing a third kind of magnetism in addition to the two long-known kinds of magnetism: ferromagnetism and antiferromagnetism. Altermagnetism appear in materials due to non-relativistic spin-orbit coupling in the non-interacting electronic band structure and is thus not due to electronic interactions, usually associated with magnetism.

The unconventional mechanism behind altermagnetism also leads to completely different symmetry properties. In altermagnets, the magnetization appearing due to broken Kramer's spin-degeneracy is momentum dependent, with sign changing values and nodes. Notably, due to the sign changes, the net magnetization is still zero in an altermagnet. Altermagnetism has already been proposed to be present in many materials with majority of them displaying a $d$-wave symmetry \cite{Smejkal22}, including the parent cuprate material La$_{2}$CuO$_{4}$ \cite{Smejkal22a}. Since doped cuprate materials are intrinsic superconductors with spin-singlet $d$-wave pairing symmetry \cite{Scalapino95,Tsuei00}, this provides an alluring prospect of having $d$-wave superconductivity in altermagnets.

Almost all known superconductors are believed to be well described by Bardeen, Cooper, and Schrieffer (BCS) \cite{Bardeen57} theory, where electrons with opposite momentum $k$ and $-k$ and opposite spins $\uparrow$ and $\downarrow$ pair in a spin-singlet configuration. These spin-singlet Cooper pairs become less energetically favorable when spin-degeneracy is broken, generating a finite spin-splitting due to either the application of an external magnetic field or due to the presence of intrinsic net magnetization in the material. Thus, increasing spin-splitting eventually destroys the BCS state. Still, superconductivity has been shown to survive for even larger external magnetic fields, by instead forming Cooper pairs with a finite center-of-mass momentum, resulting in finite-momentum superconductivity, originally studied independently by Fulde-Ferrell (FF) \cite{Fulde64} and Larkin-Ovchinnikov (LO) \cite{Larkin64}. 

Altermagnets, due to the distinct momentum dependence of their magnetization with no net magnetization, have already been anticipated to provide intriguing possibilities for superconductivity \cite{Mazin22}. 
In fact, spin-singlet Cooper pairs have very recently been studied theoretically \cite{Zhang23,Wei23,Ouassou23,Beenakker23,Papaj23,Hans23} in altermagnets, but then only induced by proximity effect from external superconductors, in heterostructures enticing for spintronics applications. However, despite altermagnetism being found in parent cuprate compounds, spin-singlet superconductivity as an intrinsic quantum phase of matter has not yet been explored in altermagnets.

In this Letter we investigate intrinsic superconductivity originating from effective electron-electron attraction in $d$-wave altermagnetic metals. We find a highly sought-after finite-momentum superconducting phase in systems with spin-singlet $d$-wave superconductivity, even with no net magnetization present. However, this phase is absent in systems with spin-singlet $s$-wave superconductivity, which we explain by the unusual momentum-space magnetization. 
By also applying an external magnetic field, we also uncover a rich phase diagram resembling almost the shape of a ``Yoda-ear", with a cascade of phase transitions between zero- and finite-momentum pairing and normal state phases, occurring due to an intricate balance of the spin-split Fermi surface and superconducting condensation energy. Interestingly, we find a large region of field-induced superconductivity, where superconductivity only appear at high magnetic fields from a low-field normal phase. These results establish altermagnetism as a key material property for generating multiple exotic and uncommon superconducting behaviors.

{\it{Model, methods and parameters.}}--- To capture established altermagnetism, we consider metallic $d$-wave altermagnet with the Hamiltonian \cite{Smejkal22a}
\begin{eqnarray}
&&H_0=\sum_{k,\sigma} \left(\xi_{k}-\sigma (t_{\rm am}/2)(\cos(k_x)-\cos(k_y))+\sigma B \right) c_{k \sigma}^{\dagger} c_{k \sigma} \nonumber \\
&&+ \sum_{k,k',q} V_{k,k^{\prime}} c_{k+q \uparrow}^{\dagger} c_{-k+q \downarrow}^{\dagger}c_{-k'+q \downarrow}c_{k'+q \uparrow}, \label{eq:Hamil}
\end{eqnarray} 
where $c_{k \sigma}^{\dagger}$ ($c_{k \sigma}$) is the creation (annihilation) operator of an electron with spin $\sigma$ and momentum $k$, $\xi_{k}$ is the (spin-independent) electron band dispersion, and $t_{\rm am}$ is the strength of the $d$-wave altermagnetic spin-splitting, originating from electric crystal fields. Eq.~\eqref{eq:Hamil} encodes the two bands closest to the Fermi level, thus relevant for superconductivity, in the minimal four-band lattice model of Ref.~[\onlinecite{Smejkal22a}]. For simplicity we consider the band dispersion of a square lattice, given by $\xi_{k}=-2t(\cos(k_x)+\cos(k_y))-\mu$, with $t=1$ the nearest-neighbor hopping amplitude set as the energy unit, and $\mu$ the chemical potential tuned to fix the average density of electrons $\rho=\sum_{k,\sigma}\langle c^{\dagger}_{k\sigma}c_{k\sigma}\rangle$. We also include an in-plane external magnetic field $B$ (with the electron magnetic moment $\mu_B=1$) for controlling a Zeeman spin-splitting but with no orbital effects expected. Intrinsic superconductivity is provided by a generic effective attraction for spin-singlet pairing, $V_{k,k^{\prime}}$. We primarily consider $d$-wave superconductivity, as present in the cuprate superconductors \cite{Scalapino95,Tsuei00}, which can most simply be generated by a nearest-neighbor attraction
\begin{equation}
V_{k,k^{\prime}}=-V\left(\gamma(k)\gamma(k')+\eta(k)\eta(k')\right), \label{eq:int} 
\end{equation}
where $\gamma(k)=\cos(k_x)+\cos(k_y)$ and $\eta(k)=\cos(k_x)-\cos(k_y)$ are the two form factors for nearest-neighbor interaction on a square lattice, and $V$ is a constant attraction strength. The nodes of the considered $d$-wave superconductivity lie along the $k_x=\pm k_y$ lines and matches the chosen directions of the altermagnet nodes in Eq.~\eqref{eq:Hamil} \footnote{The choice is motivated by the similar momentum dependence of altermagnetism and superconductivity in parent and doped cuprate material La$_2$CuO$_4$, respectively \cite{Smejkal22a}.}. For comparison, we also consider conventional, isotropic, $s$-wave pairing, using $V_{k,k^{\prime}}=-V$.

We consider only spin-singlet superconductivity, since pairing in the spin-triplet channel is rare. Hence, we do a mean-field decomposition of the Hamiltonian in Eq.~\eqref{eq:Hamil} in the spin-singlet Cooper channel resulting in
\begin{eqnarray}
H_{\rm {MF}}&=&\sum_{k,\sigma} \xi_{k \sigma} c_{k \sigma}^{\dagger} c_{k \sigma} + \sum_{k} \left( \Delta^{Q}_{k} c_{-k+Q/2 \downarrow} c_{k+Q/2 \uparrow} + \textrm{H.c.} \right) \nonumber \\
&&+ \text{ constant},
\label{eq:Hamilmf}
\end{eqnarray}
where now $\xi_{k \sigma}=\xi_{k}+\sigma (t_{\rm am}/2)(\cos(k_x)-\cos(k_y))+\sigma B$ and $\Delta^{Q}_{k}$ is the spin-singlet superconducting order parameter obtained by the self-consistency relation 
\begin{equation}
\Delta^{Q}_k=\sum_{k^{\prime}}V_{k,k^{\prime}} \langle c_{k^{\prime}+Q/2 \uparrow}^{\dagger} c_{-k^{\prime}+Q/2 \downarrow}^{\dagger} \rangle,
\label{eq:sc}
\end{equation}
with $Q$ being the finite center-of-mass momentum of the Cooper pair. For $t_{\rm am} = B = 0$, only zero-momentum ($Q = 0$), or simply BCS pairing, is present, but due to the altermagnetism and a finite magnetic field, we always allow for a finite $Q$.
Here, we only consider a single $Q$ value and focus on FF phase, where the phase of the superconducting order parameter varies but the amplitude does not \footnote{We do not consider the LO phase due to additional complexity of related charge density waves \cite{Datta19}, as already the FF phase shows that finite-momentum pairing is favorable. The total current of the FF phase is still zero, despite the finite $Q$ \cite{Fulde64}.}. 
Incorporating the momentum dependence of $V_{k,k^{\prime}}$ in Eq.~\eqref{eq:int} we can write $\Delta^{Q}_k=\Delta^{Q}_d\eta(k)+\Delta^{Q}_s\gamma(k)$, with $\Delta^{Q}_d$ being the $d$-wave superconducting order parameter and $\Delta^{Q}_s$ being the extended $s$-wave superconducting order parameter \cite{SudboBook}, where both $\Delta^{Q}_{s,d}$ parametrically depending on $Q$. We solve the Hamiltonian $H_{\rm {MF}}$ in Eq.~\eqref{eq:Hamilmf} self-consistently using Eq.~\eqref{eq:sc} for fixed $Q$, and then obtain the true ground state by minimizing the ground state energy, $E=\sum_{k,\sigma}\xi_{k \sigma}\langle c^{\dagger}_{k\sigma}c_{k\sigma} \rangle-(\Delta^{Q}_d)^2/V-(\Delta^{Q}_s)^2/V+\mu\rho$, with respect to $Q$.  

In the following we report results for $V=2$ and $\rho=0.6$ on a square lattice of size $1000\times 1000$, enough to mimic the thermodynamic limit and capture relevant values of $Q$. Other values of $V$ and $\rho$ give no qualitative difference, see Supplementary Material (SM) \cite{SM}. $Q$ is a vector with two possible directions in two dimensions. We show in the SM \cite{SM} that the ground state energy minima occurs for a uniaxial $Q$ along the $x$-axis and we thus only show results for uniaxial $Q$, setting $Q_x\equiv Q$ for simplicity. We further find that $\Delta^{Q}_s$ is very small compared to $\Delta^{Q}_d$ for all investigated parameters, thus we only report values for $\Delta^{Q}_d$. 

\begin{figure}[t]
\includegraphics[width=1.0\linewidth]{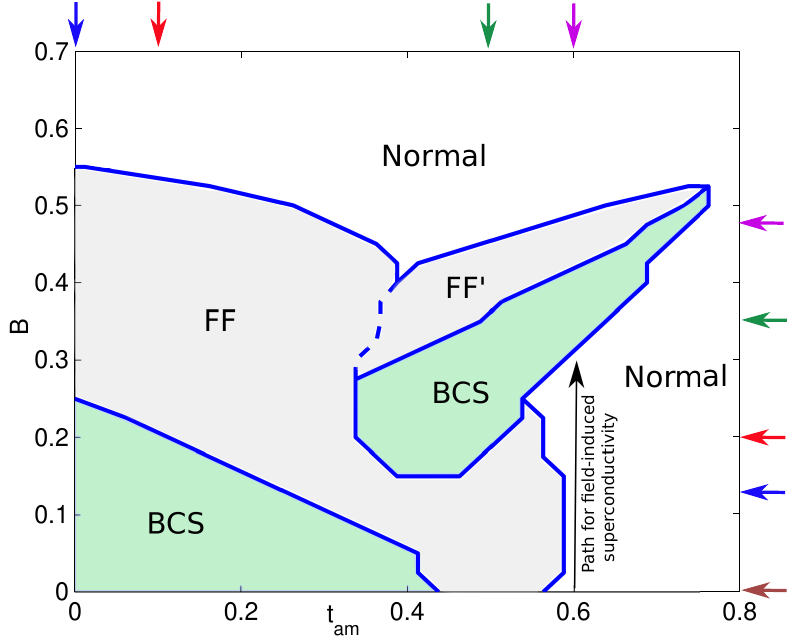} 
\caption{Phase diagram of $d$-wave superconductivity in the $B$-$t_{\rm am}$ plane indicating finite-momentum (FF) superconducting phase (gray), BCS zero-momentum superconductivity (green), and normal phase (white) with boundaries in-between (blue lines) and between different FF phases (dashed blue line).
Normal phase is identified as $\Delta^{Q}_d<0.0009$ for all $Q$ values. Calculations are performed at a set of discrete points in the $B$-$t_{\rm am}$ plane, spaced $0.025$ apart, with blue lines drawn by taking the midpoint of the two values of $t_{\rm am}$ hosting different phases for a fixed $B$.  Long arrow indicates one path for field-induced superconductivity, while short colored arrows indicate line-cuts in Fig.~\ref{fig:linecuts}.
 }
\label{fig:phasediagram} 
\end{figure}

{\it{Results.}}--- We first show in Fig.~\ref{fig:phasediagram} the ground state phase diagram obtained by varying $B$ and $t_{\rm am}$, for a range of realistic strengths \cite{Smejkal22}. The phase diagram broadly consists of three different phases: BCS phase (green), finite-momentum FF phase (gray), and normal phase with no superconductivity (white).  Here, we characterize the BCS phase as when the ground state energy for $Q=0$ is the lowest, while for the FF phase a $Q \ne 0$ has the lowest energy. For comparison we report the same phase diagram for conventional $s$-wave superconductivity in the SM \cite{SM}.

Focusing first on the situation with no applied magnetic field, $B=0$, we find a finite-momentum FF phase for  a range of finite altermagnet strengths ($0.44 \lesssim t_{\rm am} \lesssim 0.56$). This is remarkable since here the FF phase is obtained without any net magnetization in the system, in contrast to its usual occurrence in finite fields \cite{Fulde64}. This finding is perhaps even more surprising when noting that an FF phase is absent for $s$-wave superconductors at zero field, see SM \cite{SM}. 
The emergence of a $B=0$ FF phase in $d$-wave superconductors can be understood by considering the nodal structure of both the superconductor and altermagnet, see Fig.~\ref{fig:FSreentrant}(a). Since the nodes for both the altermagnet (where there is no spin-splitting) and the $d$-wave superconducting order parameter (where the order parameter is zero) lie along the $k_x=\pm k_y$ lines, the gapped parts of the Fermi surface, with finite superconducting order parameter, always host a finite magnetization due to the altermagnetism. As a result, electrons with $k, \uparrow$ only finds their spin-singlet BCS Cooper partners $-k, \downarrow$ at different energies due to the finite spin splitting, while pairing $k, \uparrow$ and $-k+Q, \downarrow$ can still occur at the zero energy difference, thus resulting in a finite $Q$ FF ground state. In contrast, $s$-wave superconductors have no superconducting nodes and the system can then still gain sufficient condensation energy by forming zero-momentum BCS pairs around the altermagnet nodal points, where the Fermi surface retains its spin degeneracy, thus preventing an FF ground state.
 
 Next, considering the $t_{\rm am}=0$ line, we find the well-established transitions of BCS to FF to normal phases with increasing $B$ \cite{Fulde64}, but with increasing $t_{\rm am}$ and finite $B \ne 0$ this drastically changes and we uncover an interesting phase diagram looking a bit like a ``Yoda-ear". For weak $t_{\rm am} \lesssim 0.34$, transitions are similar to $t_{\rm am}=0$, though with increasing $t_{\rm am}$, the critical $B$ required for the BCS to FF transition is reduced, eventually reaching zero for $t_{\rm am}\approx 0.44$ as discussed in the previous paragraph. However, for $t_{\rm am}\gtrsim 0.34$, the BCS phase re-appears at higher $B$, thus generating a cascade of phase transitions. For example, in the regime $0.44 \lesssim t_{\rm am} \lesssim 0.56$, the system shows three different transitions with increasing $B$: one from FF to BCS, next from BCS to another FF$^\prime$ phase, and eventually from FF$^\prime$ phase to normal phase. The large-$B$ FF$^\prime$ phase is characterized as a different FF phase due to a distinctly different $Q$ vector, as detailed in the next paragraph. Another remarkable feature occurs for $0.59 \lesssim t_{\rm am} \lesssim 0.76$. In this regime of altermagnetism, the system is in a normal phase at zero magnetic field $B=0$, but then superconductivity emerges with increasing magnetic field, first by forming a BCS phase and then transitioning into the FF$^\prime$ phase. In Fig.~\ref{fig:phasediagram} we illustrate one such path of field-induced superconductivity. Field-induced superconductivity should be impossible in spin-singlet superconductors, but here the altermagnetism provides a route to still generate it.

\begin{figure}[t]
\includegraphics[width=1.0\linewidth]{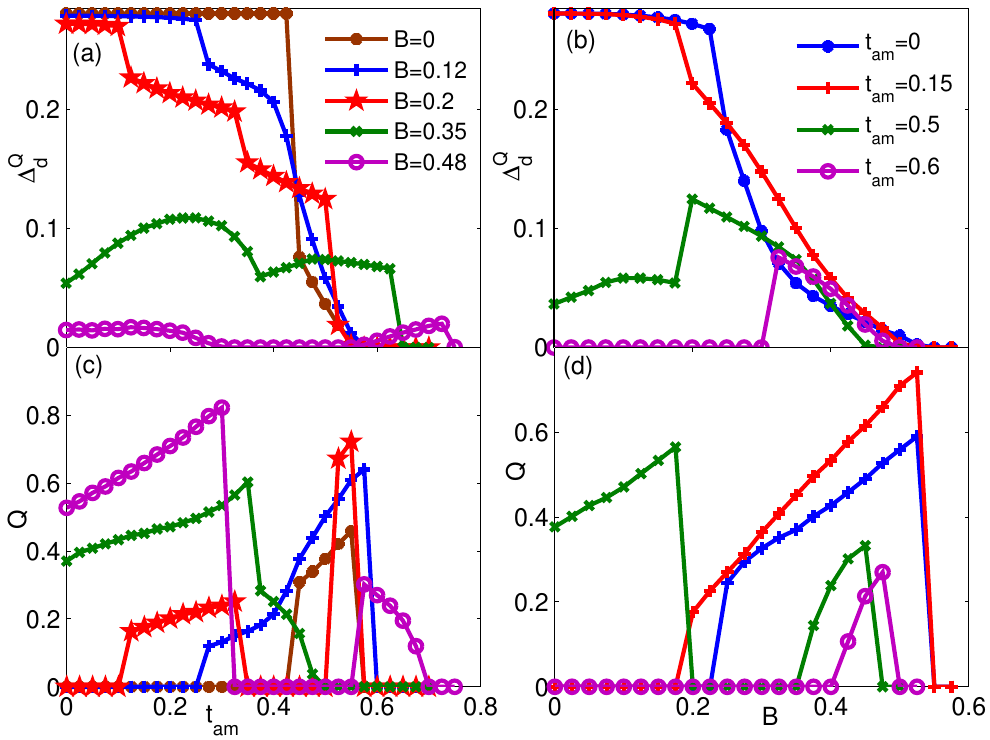} \caption{Superconducting order parameter $\Delta^{Q}_d$ (a,b) and values of $Q$ (c,d) in the different ground states (lowest total energy) at fixed values of $B$, varying $t_{\rm am}$ (a,c) and fixed values of $t_{\rm am}$, varying $B$ (b,d).}
\label{fig:linecuts} 
\end{figure}

In order to understand superconductivity within each individual phase, we show in Fig.~\ref{fig:linecuts} the superconducting order parameter $\Delta^{Q}_d$ (a,b) and  $Q$ values (c,d) in the ground state for different line cuts of the phase diagram in Fig.~\ref{fig:phasediagram}, indicated by color arrows and chosen to capture all distinct phase transitions. In Fig.~\ref{fig:linecuts}(a,c), we show five different line cuts for fixed $B$ values, varying $t_{\rm am}$. For $B=0$ (brown dot), $Q$ becomes finite in the region $0.44 \lesssim t_{\rm am} \lesssim 0.56$, capturing the FF phase even in the absence of applied field. The corresponding $\Delta^{Q}_d$ display an expected jump \cite{Fulde64} from the BCS value to a lower value in the FF phase and with further reduction toward zero with increasing $t_{\rm am}$. For a higher $B=0.12$ (blue plus), the FF phase is found over an even wider range of $t_{\rm am}$, with a notable monotonic increase in $Q$. Here the jump in $\Delta^{Q}_d$ from the BCS value to the FF value is notably suppressed. Further increasing the magnetic field to $B=0.2$ (red star) results in four transitions as $t_{\rm am}$ is increased: BCS to FF, FF to BCS, BCS to FF, and eventually FF to normal phase, with both $Q$ and $\Delta^{Q}_d$ displaying jumps between BCS to FF or FF to BCS transitions. The difference in $Q$ values of the two FF phases separated by the intermediary BCS phase can be thought of as the reminiscent of the notable monotonic increase in $Q$ in the FF phase for lower $B$ (compare blue plus and red star curves). For $B=0.35$ (green cross), the system instead goes from one FF phase directly to another FF$^\prime$ phase with increasing $t_{\rm am}$, with a distinct jump in $Q$ values at the transition. As we establish in the SM \cite{SM}, this is due to two competing FF states with different $Q \neq 0$ in this regime of $B$. The global energy minima is obtained for one $Q$ for a range of $t_{\rm am}$ and then the energy balance shifts to the other $Q$ at higher $t_{\rm am}$. With even further increase in $t_{\rm am}$, the $Q=0$ solution becomes most energetically favorable, before eventually reaching the normal phase at large $t_{\rm am}$. Due to this competition between different local energy minima, $\Delta^{Q}_d$ shows a non-monotonic behavior with increasing $t_{\rm am}$ and is also suppressed at the FF to FF$^\prime$ transition. This suppression is enhanced for larger $B$, eventually reducing $\Delta^{Q}_d$ to zero resulting in a normal phase between two FF phases, as seen for $B=0.48$ (magenta circle).

The field behavior including field-induced superconductivity is most prominent in line cuts at fixed $t_{\rm am}$ and varying $B$, as shown in Fig.~\ref{fig:linecuts}(b,d). For $t_{\rm am}=0.0,0.15$ (blue dot, red plus), $\Delta^{Q}_d$ and $Q$ shows behavior expected in a spin-singlet superconductor in an applied magnetic field, with the BCS phase giving way to the FF phase at larger fields \cite{Fulde64}. Here, larger $t_{\rm am}$ makes the FF phase occurring in a larger parameter space. For $t_{\rm am}=0.5$ (green cross) the situation is notably changed. At zero and low fields an FF phase is present, which then transitions into a BCS phase at finite fields, a transition that is accompanied with a notable large increase in $\Delta^{Q}_d$. Thus an applied magnetic field here causes a strengthening of superconductivity. Beyond the BCS phase, another FF$^\prime$ phase appear, before transitioning into the normal phase. Finally, at higher $t_{\rm am}=0.6$ (magenta circle), $\Delta^{Q}_d$ instead clearly jumps from zero to a finite value with increasing magnetic field, showing the emergence of field-induced superconductivity.

The remarkable finding of field-induced superconductivity can be understood by looking at the normal state band structures. For clarity we focus on the field-induced path marked in Fig.~\ref{fig:phasediagram} and show in Fig.~\ref{fig:FSreentrant} the normal state Fermi surface of opposite spins for increasing $B$ at fixed $t_{\rm am}=0.6$. For $B=0$ in (a), the Fermi surfaces of $\uparrow$-spin (yellow) and $\downarrow$-spin (blue) are split significantly, especially in regions away from the superconducting nodes (dashed green), implying that $k,\uparrow$ and $-k,\downarrow$ electrons are far apart in energy in these regions.
Consequently, spin-singlet pairing is energetically unfavorable, which explains the lack of zero-field superconductivity. As shown in (b), increasing the magnetic field to $B=0.38$ compensates the spin-splitting due to altermagnetism on some parts of the Fermi surface. This feature is related to a topological nodal to nodeless transition in altermagnets \cite{Fernandes24}. In particular, regions of the Fermi surface near $k_x=0$ have now almost no spin splitting and spin-singlet zero-momentum BCS pairing can thus be realized here, see the pair density plot in (d). Notably, finite $B$ makes the Fermi surfaces asymmetric between $k_y=0$ and $k_x=0$, with parts near $k_y=0$ regions still showing notable spin splitting. Now, since a $d$-wave superconducting gap has maxima in its anti-nodal regions (i.e.~around the $k_{x,y}=0$ regions), a finite condensation energy is possible by producing spin-singlet pairing near the $k_x=0$ regions. In contrast, for $s$-wave superconductors the gap is isotropic and thus the condensation energy gain with pairing only around the $k_x=0$ regions does not stabilize  $s$-wave superconductivity. This makes field-induced $s$-wave superconductivity absent in the $s$-wave superconductivity phase diagram, see SM \cite{SM}. Further increase in magnetic field to $B=0.48$ in (c) results in a separation of the two spin Fermi surfaces also in the $k_x=0$ regions. Subsequently, BCS pairing becomes unstable and finite-momentum FF$^\prime$ pairing instead occurs, before eventually, for even stronger $B$, the spin-splitting is significant for all momenta and all types of superconductivity is destroyed. 

\begin{figure}[t]
\includegraphics[width=1.0\linewidth]{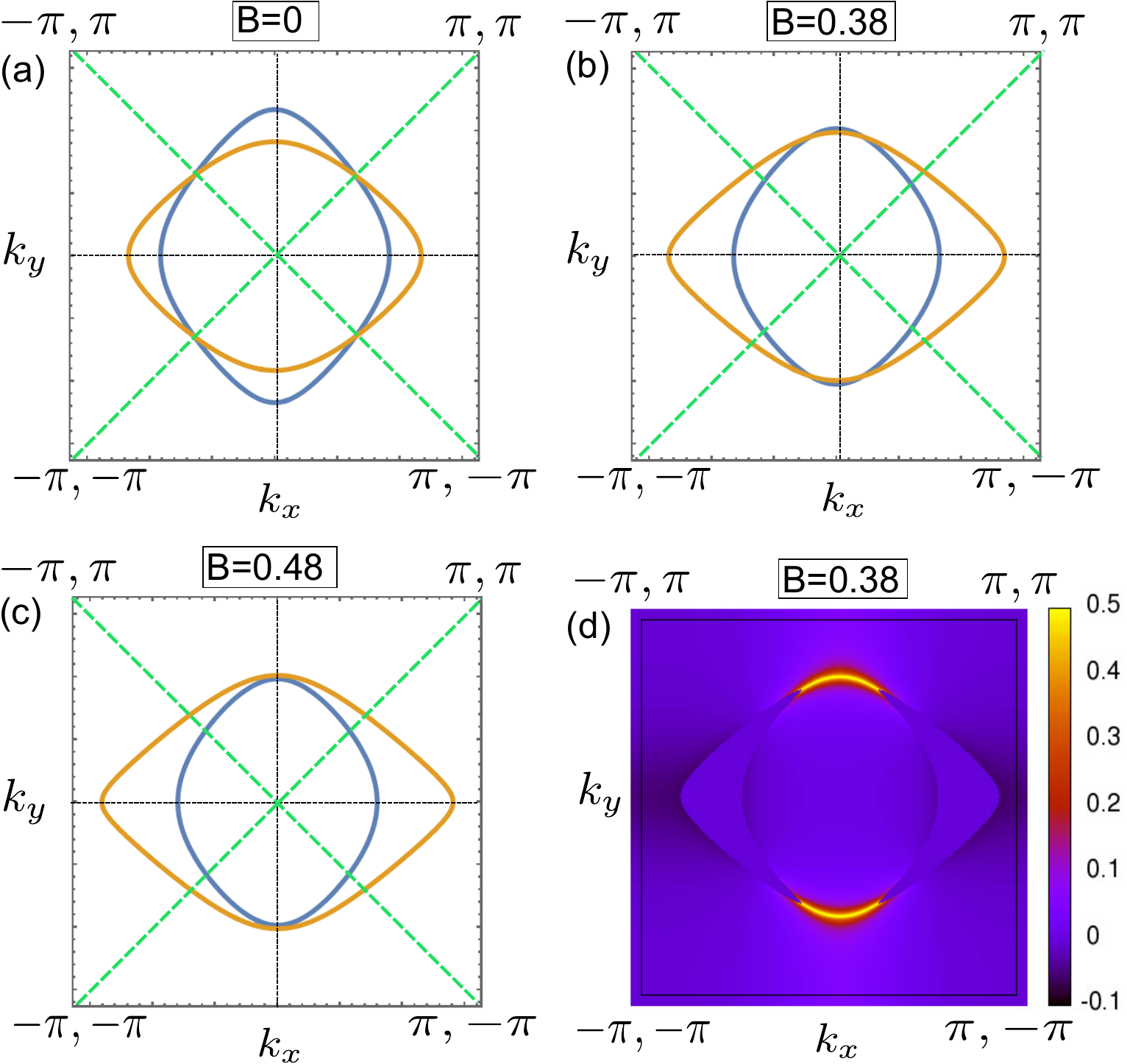} \caption{(a-c) Contour plots of the normal state electronic bands $\xi_{k \uparrow}=0$ (yellow) and $\xi_{k \downarrow}=0$ (blue) signifying the spin-split Fermi surfaces present at $t_{\rm am}=0.6$. Dashed green lines mark the $d$-wave superconducting order parameter nodes. (d) Pair density $\langle c^{\dagger}_{k}c^{\dagger}_{-k}\rangle$. }
\label{fig:FSreentrant} 
\end{figure}

{\it{Discussion.}}---Considering spin-singlet superconductivity in altermagnets we find a finite-momentum (FF) superconducting phase in the absence of an applied magnetic field. This zero-field FF phase is dependent on coinciding superconducting gap and altermagnet nodes, and do not appear in an $s$-wave superconductor. In the presence of external magnetic field, we also uncover field-induced superconductivity, due to an intricate interplay between Fermi surface shape and superconducting condensation energy. Although we are primarily concerned with $d$-wave symmetry, our results are also directly applicable to $g$- and $i$-wave altermagnets, as long as the superconducting pairing at least partially has the same nodal structure, see SM \cite{SM}. We further validate that our finding of a zero-field FF phase is robust and present also when altermagnetism is induced by interactions \cite{Maier23}, see SM.

Our finding of zero-field finite-momentum superconductivity is remarkable since the net magnetization is always zero in altermagnets. Finding finite-momentum superconductivity in microscopic models in the absence of applied magnetic field has been a long-standing unsolved theory problem \cite{Chen23,Wu23,Huang22,Agterberg20}, despite mounting experimental evidence of such a phase, often referred to as pair density waves \cite{Hamidian16,Edkins19,Liu21,Kasahara20,Chen21,Gu23,Liu23,Zhao23}. Our work provides one straight-forward path to realize finite-momentum superconductivity likely applicable to many materials, including the possibility of being the origin of such a phase in the cuprate superconductors, without needing preceding charge density order \cite{Edkins19,Agterberg20,Dai18,Dai20,Lee14,Norman18,Chakraborty19}. Moreover, finite-momentum superconductivity may be important for technological applications, as e.g.~illustrated by superconducting diode effects \cite{Ando20,Daido22,Yuan22}, opening up for large technological potential for superconducting altermagnets. 
Another remarkable feature of our results is the presence of a field-induced superconducting phase at experimentally accessible fields, see SM \cite{SM}. Such a phase is both rare \cite{Meul84,Uji01,Konoike04} and unexpected, but has gained renewed interest after its recent observation in UTe$_2$ \cite{Ran19}. 
Our work also opens for many other exotic possibilities of superconductivity. Investigating the relation of the finite-momentum Cooper pairs in altermagnets and odd-frequency pairing, or generally finite-energy pairs, is one interesting direction \cite{Chakraborty21, Chakraborty22a, Chakraborty22b}. Additional presence of relativistic spin-orbit coupling can further provide a platform for studying the interplay of finite-momentum and topological superconductivity \cite{Zhu23,Ghorashi23}. 

Although superconductivity has not yet been discovered experimentally in altermagnetic materials within the two years of their discovery, our results point out several promising directions. With too large altermagnetic spin-splitting only the normal phase is reached and thus limiting altermagnetism is favorable. Or, for strong altermagnetism, applying an external magnetic field, can be used to induce superconductivity. Moreover, $d$-wave superconductivity is likely much more amenable to altermagnetism than conventional $s$-wave superconductivity.
Alternatively, doping altermagnet insulators, making them metallic, is also a promising route.

Note added: After submission we became aware of Refs.~[\onlinecite{Rodrigo14}] and [\onlinecite{Shuntaro23}], where a zero-field finite-momentum phase is discussed in the context of cuprates with nematic-spin-nematic order and organic conductors with antiferromagnetism, respectively.

\begin{acknowledgments}
{\it{Acknowledgments.}}--- We gratefully acknowledge financial support from the Knut and Alice Wallenberg Foundation through the Wallenberg Academy Fellows program the Swedish Research Council (Vetenskapsr\aa det grant agreement no.~2018-03488) The computations were enabled by resources provided by the National Academic Infrastructure for Supercomputing in Sweden (NAISS) at the Uppsala Multidisciplinary Center for Advanced Computational Science (UPPMAX) funded by the Swedish Research Council through grant agreement no.~2022-06725.
\end{acknowledgments}

 \bibliographystyle{apsrev4-1}
\bibliography{Cuprates}

\pagebreak
\widetext
\clearpage 
\normalsize
~\vspace{0.2cm} 
\setcounter{equation}{0}
\setcounter{figure}{0}
\setcounter{table}{0}
\setcounter{page}{1}
\makeatletter
\renewcommand{\theequation}{S\arabic{equation}}
\renewcommand{\thefigure}{S\arabic{figure}}
\renewcommand{\figurename}{FIG.}

\begin{center}
{\large\bf Supplementary Material for \\
 ``Zero-field finite-momentum and field-induced superconductivity in altermagnets"}
\end{center}

\vspace{0.2cm}

In this Supplementary Material (SM), we provide additional information to support our results and conclusions in the main text. In particular, in order, we provide the phase diagram for $s$-wave superconductivity in the presence of an applied magnetic field in altermagnets for the same parameters as used for $d$-wave superconductivity in the main text. We then display the ground state energy of the $d$-wave superconductor treated in the main text in the full $(Q_x,Q_y)$-space for two representative $B,t_{\rm am}$ values and provide line cuts for varying $t_{\rm am}$ for fixed $B=0.35$ in order to illustrate phase transitions between different $Q$-values. In addition, we show the phase diagram for smaller nearest-neighbor attractive interaction strength than in the main text, and, finally, we show the robustness of our findings using a very different model of altermagnetism based on altermagnetism generated by electronic interaction.

\subsection{Phase diagram for $s$-wave superconductivity}
In Fig.~1 of the main text we show the phase diagram of $d$-wave superconductivity in the simultaneous presence of altermagnetism and external magnetic field. Here, we report the corresponding phase diagram for $s$-wave superconductivity. We consider an onsite effective electron-electron attraction, resulting in the interaction $V_{k,k^{\prime}}=-V$, i.e.~no momentum-dependent form factors, in contrast to Eq.~(2) of the main text for $d$-wave superconductivity. Such an attraction leads to conventional  spin-singlet $s$-wave superconductivity, with the superconducting order parameter $\Delta^{Q}_0$ no longer momentum-dependent and thus fully isotropic in momentum space. We again solve the self-consistent gap equation following the same procedure and using the same parameters as in the main text, but now with $V_{k,k^{\prime}}=-V$ and $\Delta^{Q}_k=\Delta^{Q}_0$ in Eqs.~(1-4) of the main text. 

\begin{figure}[htb]
\includegraphics[width=0.5\linewidth]{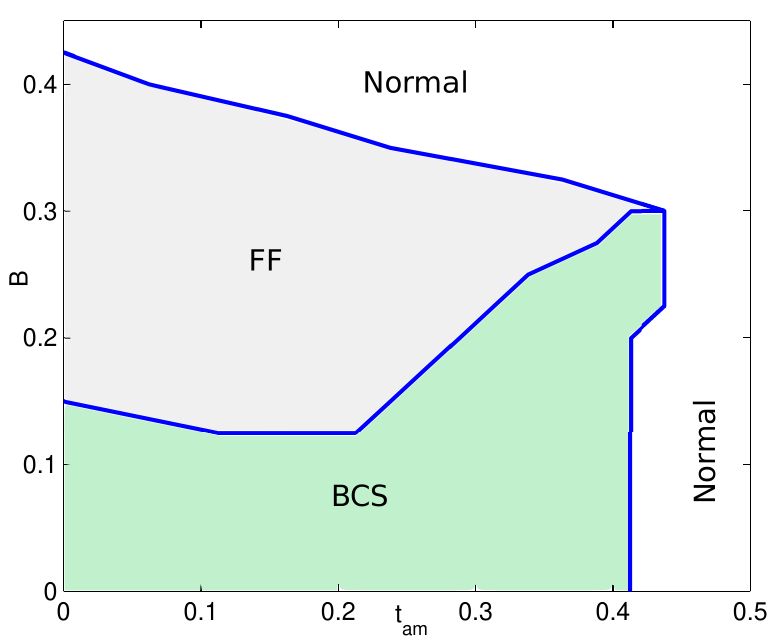} \caption{Phase diagram of $s$-wave superconductivity in the $B$-$t_{\rm am}$ plane, indicating finite-momentum (FF) superconducting phase (gray), BCS zero-momentum superconductivity (green), and normal phase (white) with boundaries in-between (blue lines) (same as Fig.~1 in the main text but instead for $s$-wave superconductivity).
Normal phase is identified as $\Delta^{Q}_0<0.0009$ for all $Q$ values. Calculations are performed at a set of discrete points in the $B$-$t_{\rm am}$ plane, spaced $0.025$ apart, with blue lines drawn by taking the midpoint of the two values of $t_{\rm am}$ hosting different phases for a fixed $B$. 
}
\label{fig:swavePD} 
\end{figure}

In Fig.~\ref{fig:swavePD} we show the resulting phase diagram in the $B$-$t_{\rm am}$ plane. As clearly seen, we do not find a finite-momentum FF state for $B=0$. The reason for not finding the FF state at zero field is the isotropic nature of $s$-wave superconductors. The electrons near the nodes of the altermagnet do not break the spin-degeneracy. As a result, the condensation energy gain due to BCS ($Q=0$) pairing near the altermagnetic nodal regions prohibits the system to form any finite-momentum superconducting pairing. This is in contrast to $d$-wave superconductors where the altermagnetic nodes coincides with the nodes of the superconductor, which essentially eliminates pairing at the altermagnet nodes. With increasing $B$, but staying at $t_{\rm am}=0$, the FF phase appears, as is well-known \cite{Fulde64}. This FF phase persists for finite $t_{\rm am}$, but the FF region shrinks for larger $t_{\rm am}$. Eventually, at large $B$ and $t_{\rm am}$, the system goes into a normal phase with no superconductivity. The critical $t_{\rm am}$ for the BCS to normal transition increases only slightly for high $B$, this essentially eliminates the region of field-induced superconductivity that is clearly present for $d$-wave superconductivity (although for some parameters small regions can still exist). Again, it is the lack of superconducting nodes in $s$-wave superconductors, that makes the interesting ``Yoda-ear" phase diagram obtained for $d$-wave superconductors disappear here.     

\subsection{Ground state energy as a function of $(Q_x,Q_y)$}

\begin{figure*}[htb]
\includegraphics[width=0.7\linewidth]{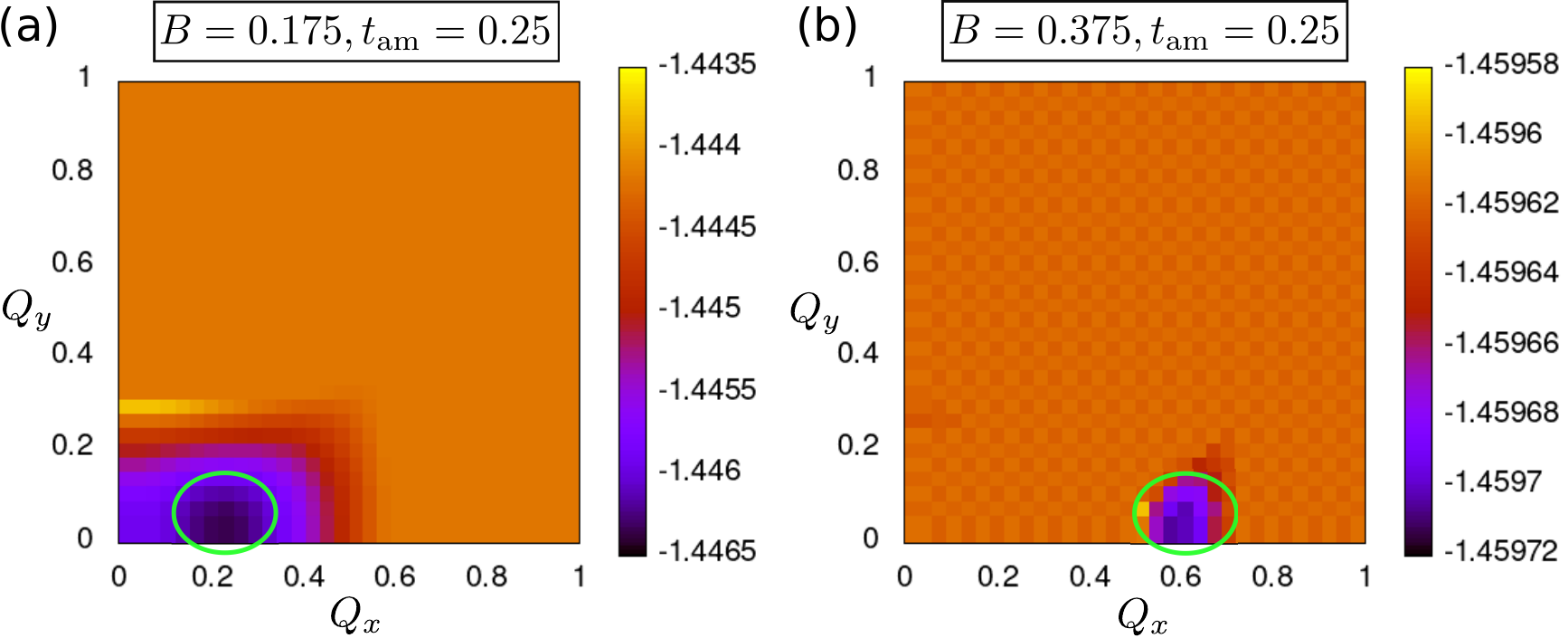} \caption{Ground state energy E for $d$-wave superconductivity as a function of ($Q_x,Q_y$) for $t_{\rm am} = 0.25$ and $B = 0.175$ (a) and $B = 0.375$ (b). Green circles indicate the energy minimum. For computational reasons, results are obtained for a smaller lattice of $200\times 200$.}
\label{fig:fullQ} 
\end{figure*}
In the main text, we only discuss uniaxial $Q$ vectors, using the notation $Q=(Q_x,0)$. This choice is motivated by our findings of the ground state energy in the full momentum space $(Q_x,Q_y)$. In Fig.~\ref{fig:fullQ}, we demonstrate these results by plotting the ground state energy $E$ in the full $(Q_x,Q_y)$-space for two representative $B$ and $t_{\rm am}$ values. We check for other $B,t_{\rm am}$ values and find similar features. As seen in Fig.~\ref{fig:fullQ}, the ground state energy minima clearly occur on the $Q_y=0$ line. Note that rotational symmetry is broken due to the presence of a finite $B$, as it is already evident from the Fermi surfaces shown in Fig.~3 of the main text. For $B=0$ (not shown), the rotational symmetry is preserved and the energy minimum then occurs on both the $Q_y=0$ and $Q_x=0$ lines, at the same $Q_x$/$Q_y$ values. These results establish that using only uniaxial $Q=(Q_x,0)$ is sufficient.

\subsection{Phase transitions between different center-of-mass momenta $Q$}
\begin{figure*}[htb]
\includegraphics[width=1.0\linewidth]{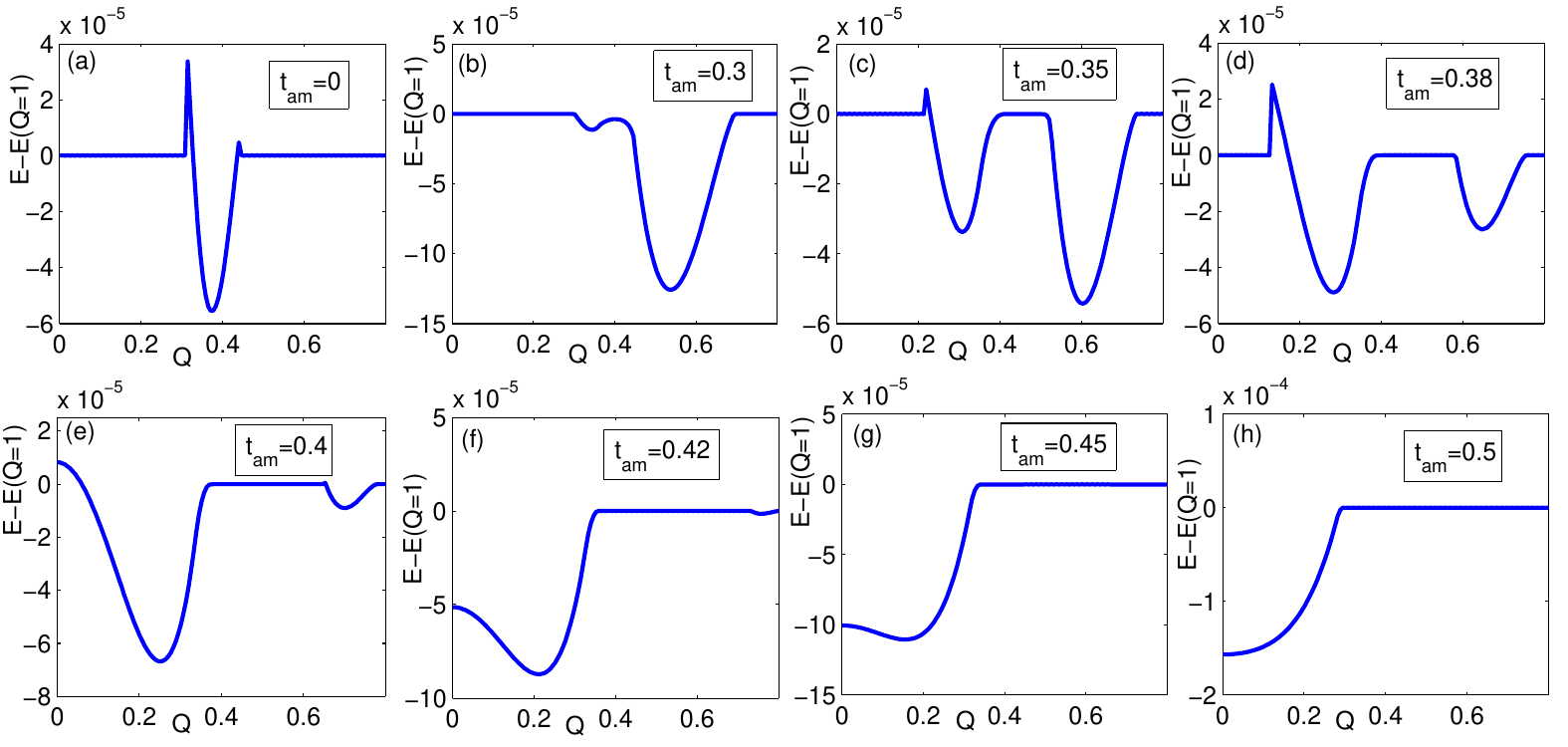} \caption{Ground state energy E for $d$-wave superconductivity as a function of $Q = Q_x$ for a fixed $B=0.35$ and increasing $t_{\rm am}$. For visualization purposes we subtract $E$ at $Q=1$ where $\Delta_{d}^{Q}=0$.}
\label{fig:energy} 
\end{figure*}

The phase diagram in Fig.~1 of the main text is obtained by analyzing the ground state energy $E$ as a function of varying $Q$. The richness of the phase diagram is due to the energy landscape and its development of multiple local minima as a function of $Q$. In Fig.~\ref{fig:energy} we illustrate this phenomenon by plotting the ground state energy as a function of $Q \equiv Q_x$ for $B=0.35$ in order to capture several distinct phases seen in Fig.~1 in the main text and directly comparable to the green curves in Fig.~2(a,c) in the main text. 

Starting at $t_{\rm am}=0$ (a), $E$ has a unique minima near $Q=0.4$. With increasing $t_{\rm am}$, this minima shifts to higher $Q$ values (b,c). However, from $t_{\rm am}\approx 0.3$ (b) a new minima additionally appears at a lower, still finite, $Q$ value. The energy for this newly appearing minima eventually becomes equivalent to the minima at the higher $Q$ (c,d) and then even becomes the new global minima with increasing $t_{\rm am}$ (d). This sudden change in $Q$ is evident in the green curves in Fig.~2(a,c) in the main text near $t_{\rm am}=0.38$. We also indicate this transition as a dashed line in Fig.~1 in the main text. Further increasing $t_{\rm am}$ causes the minima at the higher $Q$ to disappear (e,f). For these $t_{\rm am}$ strengths, the energy for $Q=0$ also becomes lower and then becoming comparable to the energy of the $Q\ne 0$ minima (g), eventually making the BCS $Q=0$ state the global energy minima (h) for even larger $t_{\rm am}$, thus making BCS superconductivity appear. Finally, increasing $t_{\rm am}$ even further gives no non-zero solution of the superconducting gap equation, resulting into a transition to the normal phase. 

\subsection{Phase diagram of $d$-wave superconductivity with smaller interaction strength}
\begin{figure*}[htb]
\includegraphics[width=0.5\linewidth]{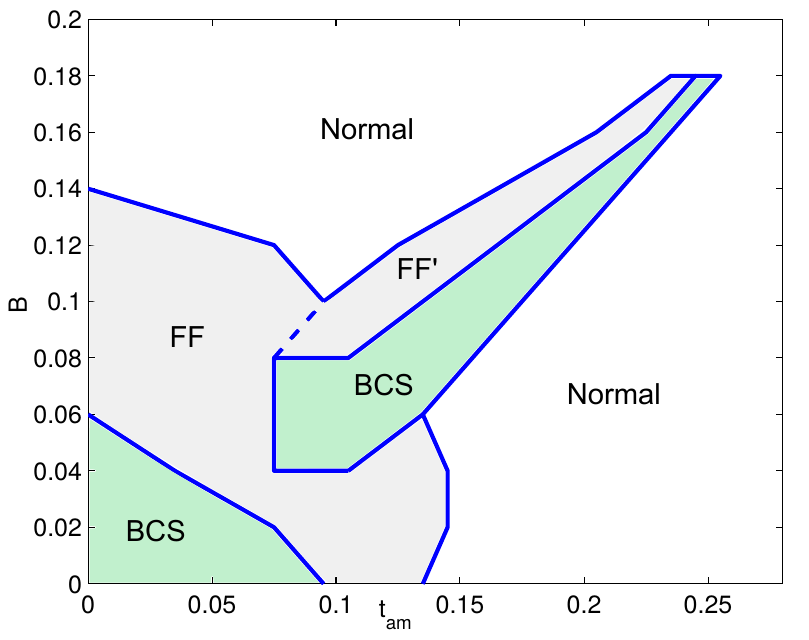} \caption{Phase diagram of $d$-wave superconductivity in the $B$-$t_{\rm am}$ plane indicating finite-momentum (FF) superconducting phase (gray), BCS zero-momentum superconductivity (green), and normal phase (white) with boundaries in-between (blue lines) and between different FF phases (dashed blue line), for a smaller interaction strength $V=1.3$ than in the main text (there $V = 2$). Normal phase is identified as $\Delta^{Q}_d<0.0002$ for all $Q$ values. Calculations are performed at a set of discrete points in the $B$-$t_{\rm am}$ plane, spaced $0.02$ apart, with blue lines drawn by taking the midpoint of the two values of $t_{\rm am}$ hosting different phases for a fixed $B$.}
\label{fig:reducedV} 
\end{figure*}

In Fig.~1 of the main text we show the ground state phase diagram of $d$-wave superconductivity with the attractive nearest-neighbor interaction strength $V=2$. Here in Fig.~\ref{fig:reducedV} we complement those results by showing the phase diagram using a smaller interaction strength, $V=1.3$. Remarkably, the phase diagram is qualiatitively very similar with a ``yoda-ear" also for $V=1.3$. This establish that our results are stable with regards to the strength of superconductivity. We however note that the scale on each axis changes, such the critical fields are now reduced. This is to be expected since reducing $V$ leads to a reduced value of $\Delta_d$ and hence the overall energy scale for superconductivity.

In Fig.~\ref{fig:reducedV} we see that field-induced superconductivity starts occurring for fields as low as $0.06$. With further lowering $V$, the magnetic fields required to have field-induced superconductivity can be made even lower (not shown). In order to get an estimate about the magnitude of fields necessary we note that the magnetic fields in Fig.~\ref{fig:reducedV} and Fig.~1 of the main text are in units of $\mu_{\rm B}/t$. We can then make an estimate of the magnetic field in Tesla. Taking a typical value of $t=100$ ~meV and with $\mu_{\rm B}=5.8 \times 10^{-2}$~meV/Tesla, a field value of $0.01$ in units of $\mu_{\rm B}/t$ is approximately $16$ Tesla. Magnetic fields of 16 Tesla, and multiples thereof, are easily accessible in high-magnetic field experiments often used for superconductors. It is therefore realistic to assume field-induced superconductivity can be achieved in altermagnets.

\subsection{Interaction-induced model for altermagnetism}

In the main text, we show results for an altermagnetic model where the band structure alone gives $d$-wave spin-splitting, i.e.~altermagnetism, due to the electric crystal potential. This is the original proposal for altermagnetism. There also exists now other proposals of generating altermagnetism. In particular, it has very recently been shown that electronic interaction combined with band structure effects can generate altermagnetism \cite{Maier23}. In order to illustrate the robustness of our results to different models of altermagnetism, we adapt in this section this interaction-induced altermagnetic model and investigate the nature of superconductivity.

Following Ref.~[\onlinecite{Maier23}], we consider a square lattice with two sublattices with the non-interacting, kinetic energy, given in momentum space as
\begin{equation}
\tilde{H}_{0}=\sum_{k,\sigma} \xi_{AA}(k) c_{A k \sigma}^{\dagger} c_{A k \sigma} + \xi_{BB}(k) c_{B k \sigma}^{\dagger} c_{B k \sigma} + \epsilon_{AB}(k) \left( c_{A k \sigma}^{\dagger} c_{B k \sigma} + h.c. \right),
\label{eq:nonint}
\end{equation}
where $c_{A/B k \sigma}^{\dagger}$ ($c_{A/B k \sigma}$) are the creation (annihilation) operators of an electron with spin $\sigma$, momentum $k$ on the sublattice $A/B$, with
\begin{eqnarray}
\xi_{AA}(k)=-2t_1\cos(2k_x)-2t_2\cos(2k_y)-\mu, \nonumber \\
\xi_{BB}(k)=-2t_2\cos(2k_x)-2t_1\cos(2k_y)-\mu, \nonumber \\
\epsilon_{AB}(k)=-2t(\cos(k_x)+\cos(k_y)).
\label{eq:dis}
\end{eqnarray}
Here $t$ is the nearest-neighbor hopping amplitude, $t_{1/2}$ are third-nearest-neighbor hopping amplitudes, and $\mu$ is the chemical potential tuned to fix the average density of electrons. We here also use the tilde-symbol to indicate that this is an alternative, interaction-induced, model for altermagnetism. 
We then add two kinds of interactions. First, an onsite repulsive Hubbard interaction is added,
\begin{equation}
\tilde{H}_{U}=\sum_{i} Un_{i\uparrow}n_{i\downarrow},
\label{eq:HUbb}
\end{equation}
where $n_{i \sigma}$ is the number operator and $U$ is the onsite coulomb repulsion. A mean-field decoupling of the Hubbard interaction in the Hartree channel gives
\begin{equation}
\tilde{H}_{U,\rm{MF}}=\sum_{i \sigma}\frac{1}{2}U\left( \rho n_{i \sigma} - m_i \sigma n_{i \sigma} \right) + \text{constant},
\label{eq:HUbbmf}
\end{equation}
where $\rho$ is the average electron density per sublattice and $m_{i}=\langle n_{i\uparrow}- n_{i\downarrow}\rangle$ is the magnetization. A homogeneous solution of the form $m_{i}=m_{A}=-m_{B}\equiv m$ gives a sublattice dependence of the magnetization, and, if non-zero, supports either antiferromagnetism or altermagnetism. As shown in Ref.~[\onlinecite{Maier23}], altermagnetism appears when there is a sublattice asymmetry with $\xi_{AA}(k) \ne \xi_{BB}(k)$. From Eq.~\eqref{eq:dis}, the condition $\xi_{AA}(k) \ne \xi_{BB}(k)$ is equivalent to $t_1 \ne t_2$. Next, in order to generate $d$-wave superconductivity, as in the main text, we also add a nearest-neighbor attractive interaction,
\begin{equation}
\tilde{H}_{V}=\sum_{\langle ij \rangle\sigma\sigma'} V n_{i\sigma}n_{j\sigma'},
\label{eq:nnscterm}
\end{equation} 
where $V$ is the nearest-neighbor attractive strength. Upon mean-field decoupling in the spin-singlet Cooper channel this term can be written in the same form as Eq.~(3) of the main text,
\begin{eqnarray}
\tilde{H}_{V,\rm{MF}}&=& \sum_{k} \left( \Delta^{Q}_{k} c_{A -k+Q/2 \downarrow} c_{B k+Q/2 \uparrow} + \textrm{H.c.} \right) \nonumber \\
&&+ \text{ constant},
\label{eq:nnsctermmf}
\end{eqnarray}
where $\Delta^{Q}_{k}$ is the spin-singlet superconducting order parameter obtained from the self-consistency relation 
\begin{equation}
\Delta^{Q}_k=\sum_{k^{\prime}}\frac{1}{2}V_{k,k^{\prime}} \langle c_{A k^{\prime}+Q/2 \uparrow}^{\dagger} c_{B -k^{\prime}+Q/2 \downarrow}^{\dagger} +h.c. \rangle,
\label{eq:scsc}
\end{equation}
with $Q$ being the finite center-of-mass momentum of the Cooper pair and the interaction $V_{k,k^{\prime}}$ given by
\begin{equation}
V_{k,k^{\prime}}=-V\left(\gamma(k)\gamma(k')+\eta(k)\eta(k')\right), \label{eq:int2} 
\end{equation}
which is the same as Eq.~(2) of the main text.

The total Hamiltonian, $\tilde{H}=\tilde{H}_0+\tilde{H}_{U,\rm{MF}} + \tilde{H}_{V,\rm{MF}}$ can be written in a matrix form using the basis $\Psi^{\dagger}=\left(c_{A k+Q/2 \uparrow}^{\dagger},c_{B k+Q/2 \uparrow}^{\dagger},c_{A -k+Q/2 \downarrow},c_{B -k+Q/2 \downarrow}\right)$ as
\begin{equation}
\tilde{H}=\sum_{k} \Psi^{\dagger} \hat{H} \Psi+\text{constant},
\end{equation}
with
\begin{equation}
\hat{H}=\left(\begin{array}{cccc} \xi_{AA} (k+Q/2)-M & \epsilon_{AB} (k+Q/2) & 0 & \Delta^{Q}_{k} \\
\epsilon_{AB} (k+Q/2) & \xi_{BB} (k+Q/2)+M  & \Delta^{Q}_{k} & 0 \\
0 & \Delta^{Q}_{k} & -\xi_{AA} (-k+Q/2)-M & -\epsilon_{AB} (-k+Q/2) \\
\Delta^{Q}_{k} & 0 & -\epsilon_{AB} (-k+Q/2) & -\xi_{BB} (-k+Q/2)+M \\
\end{array}\right).
\label{eq:Hamilmat}
\end{equation} 
Here we define $M=\frac{1}{2}Um$, while the spin-independent Hartree shift term $\frac{1}{2}U\rho$ of Eq.~\eqref{eq:HUbbmf} is absorbed in $\mu$. We then follow the same procedure as in the main text to find the ground state by self-consistently solving for all the order parameters and find the minima in the energy by varying $Q$. In addition to $\Delta^{Q}_{s,d}$ as in the main text, here $M$ is another order parameter which must also be obtained self-consistently. In the following, we report results for average density per sublattice $\rho=0.9$, $V=2$, using $t=1$ as the energy unit and the same system size as in the main text.

\begin{figure*}[htb]
\includegraphics[width=1.0\linewidth]{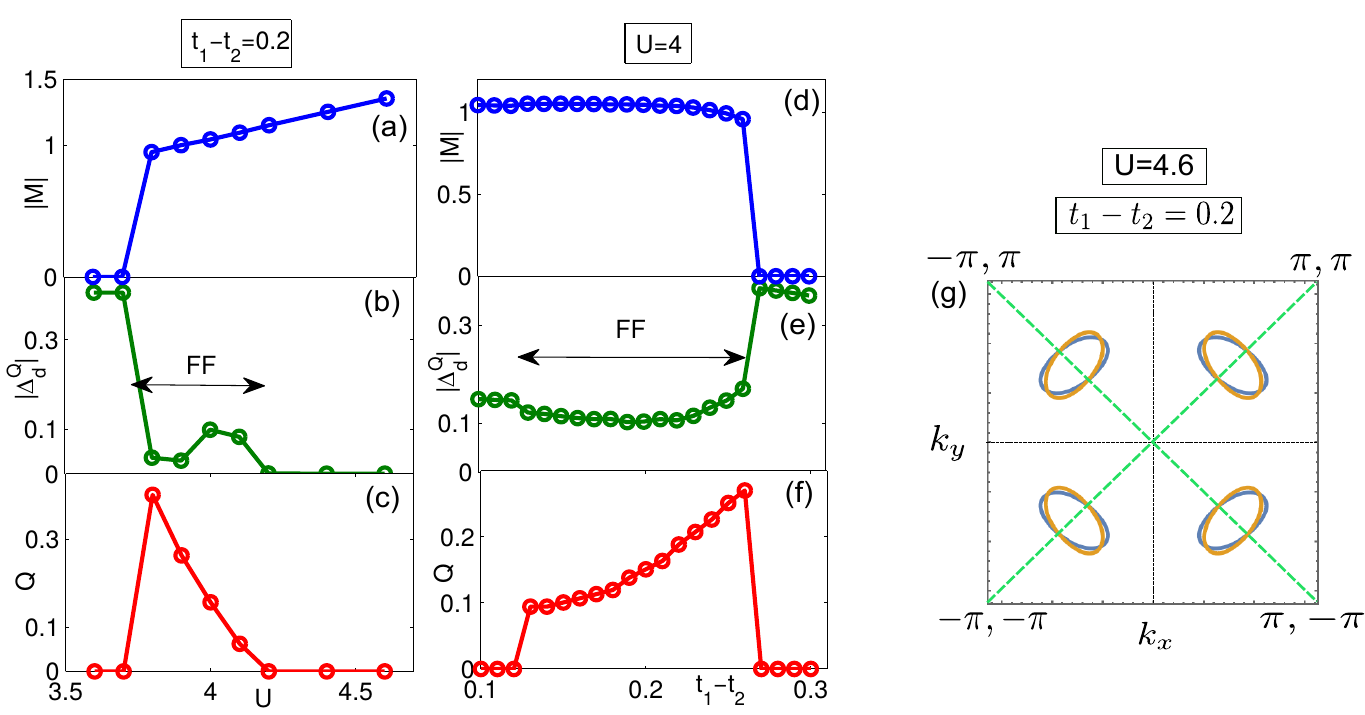} \caption{Magnetization per sublattice $|M|$ (a,d), superconducting order parameter $\Delta^{Q}_{d}$ (b,e), and values of $Q$ (c,f) in the ground states (lowest total energy) at fixed $t_1-t_2$, varying $U$ (a-c) and fixed $U$, varying $t_1-t_2$ (d-f). (g) Normal-state Fermi surface of the $\uparrow$- (yellow) and $\downarrow$-spins (blue) showing interaction-generated altermagnetic spin-splitting. Dashed green lines mark the $d$-wave superconducting order parameter nodes.}
\label{fig:intermodel} 
\end{figure*}

We first fix $t_1-t_2=0.2$ and show $|M|$, $\Delta^{Q}_d$, and $Q$ in Fig.~\ref{fig:intermodel} (a-c) with varying $U$. For $U\le 3.7$, there is no magnetization, $|M|=0$. In this regime of $U$, uniform BCS $d$-wave superconductivity exists with $Q=0$. Note that the extended $s$-wave order parameter $\Delta^{Q}_{s}$ is very small compared to $\Delta^{Q}_{d}$ and hence is not shown. At a critical $U=3.8$, $|M|$ acquires a finite large value and then increases linearly with further increase of $U$. The corresponding Fermi surfaces are shown in (g) for $U=4.6$. Due to finite $|M|$ and asymmetry $t_1-t_2$, the Fermi surfaces of $\uparrow$-spin (yellow) and $\downarrow$-spin (blue) are split, except along $|k_x|=|k_y|$ and $|k_x\pm k_y|=\pi$, with the former also present for the type of altermagnetism in the main text. This shows that a finite interaction $U$ generates altermagnetism with nodes along $|k_x|=|k_y|$ and $|k_x\pm k_y|=\pi$. Even though the nodes of the $d$-wave superconductivity (green dashed lines in (g)) now match partially with the nodes of this type of interaction-induced altermagnetism, we find a FF phase for a range of $3.7<U<4.4$, see (b,c). In order to analyze the robustness of this FF phase, we next fix a value of $U=4$ in this regime of FF phase and instead vary the asymmetry $t_1-t_2$. The obtained results are shown in Fig.~\ref{fig:intermodel}(d-f). For $t_1-t_2=0$ (not shown), a finite $|M|$ generates antiferromagnetism and not altermagnetism, with the $\uparrow$-spin and $\downarrow$-spin Fermi surfaces not split (not shown). Consequently, we find uniform BCS $d$-wave superconductivity for $t_1-t_2 \le 0.12$ due to no or low spin-splitting. However, the corresponding $|\Delta^{Q}_{d}|$ is lower than $|\Delta^{Q}_{d}|$ when $|M|=0$, compare (b) and (e). Such reduction in $|\Delta^{Q}_{d}|$ is due to the competition between superconductivity and antiferromagnetism. With increasing $t_1-t_2$, the spin-splitting increases and the FF phase appears. The FF phase survives for the range of $0.12<t_1-t_2<0.27$. For larger $t_1-t_2 \ge 0.27$, $|M|$ becomes zero since for these values of $t_1-t_2$ the critical $U$ needed to obtain finite $|M|$ is larger than $U=4$ used. With $|M|=0$, uniform BCS $d$-wave superconductivity then re-appears, and now with a sudden increase in $|\Delta^{Q}_{d}|$, due to lack of competition between superconductivity and magnetism. These results establish that zero-field FF phase is present and exists in a large range of parameters also for interaction-induced altermagnetism. The alignment of the altermagnet and superconducting nodes is a key component, just as for the altermagnetism driven by band structure effects studied in the main text.

Before ending, we note that we here do not consider an external magnetic field, but leave it for future work. We do this because it has long been known that magnetic field can induce a spin-flop transition in antiferromagnets \cite{Nagamiya55,Bogdanov07}, where spins change their alignment in a non-trivial way. It would be interesting to see if the possibility of a spin-flop transition also exists in altermagnets and establish the associated changes to superconductivity. Given this possibility, different alignments of magnetization need to be considered and hence this go beyond the scope of this current work. We emphasize that the model in the main text, i.e., a model where altermagnetism is originated from electric crystal field potential, is not going to be affected by any spin-flop transition with magnetic field since there it is the lattice structure which solely decides magnetic structure and not interaction.

In summary, the results in this section show that a zero-field FF phase is also obtained in a model where altermagnetism is generated by electronic interactions. This demonstrates that our findings in the main text are robust to different modeling schemes of altermagnetism.

\end{document}